\documentclass{article}
\usepackage{PRIMEarxiv}
\usepackage[utf8]{inputenc} 
\usepackage[T1]{fontenc}    
\usepackage{hyperref}       
\usepackage{url}            
\usepackage{booktabs}       
\usepackage{amsfonts}       
\usepackage{nicefrac}       
\usepackage{microtype}      
\usepackage{lipsum}
\usepackage{fancyhdr}       
\usepackage{graphicx}       
\graphicspath{{media/}}     
\usepackage{natbib}
\usepackage{amsfonts,amsmath,amssymb}
\usepackage[table]{xcolor}

\usepackage{subfigure,bm,yfonts,cleveref,xcolor}

\title{Deep learning unflooding\\for robust subsalt waveform inversion}

\author{
  Abdullah Alali \\
  KAUST \\
  Saudi Arabia, Thuwal \\
  \texttt{abdullah.alali.1@kaust.edu.sa} \\
  \And
  Vladimir Kazei \\
  Aramco Americas \\
  United States, Houston\\
  \texttt{vkazei@gmail.com } \\
  \AND 
\hspace{45pt}  Mahesh Kalita \\
\hspace{45pt}  KAUST, currently Shearwater \\
\hspace{45pt}  United Kingdom \\
\hspace{45pt}  \texttt{mkalita347@gmail.com } \\
  \And
  Tariq Alkhalifah \\
  KAUST \\
  Saudi Arabia, Thuwal \\
  \texttt{tariq.alkhalifah@kaust.edu.sa} \\
}

\newcommand{\norm}[1]{\left\lVert#1\right\rVert}

\begin{document}
\maketitle

\begin{abstract}
Full-waveform inversion (FWI), a popular technique that promises high-resolution models, has helped in improving the salt definition in inverted velocity models. The success of the inversion relies heavily on having prior knowledge of the salt, and using advanced acquisition technology with long offsets and low frequencies. Salt bodies are often constructed by recursively picking the top and bottom of the salt from seismic images corresponding to tomography models, combined with flooding techniques. The process is time-consuming and highly prone to error, especially in picking the bottom of the salt (\textbf{BoS}). Many studies suggest performing FWI with long offsets and low frequencies after constructing the salt bodies to correct the miss-interpreted boundaries. Here, we focus on detecting the \textbf{BoS} automatically by utilizing deep learning tools. We specifically generate many random 1D models, containing or free of salt bodies, and calculate the corresponding shot gathers. We then apply FWI starting with salt flooded versions of those models, and the results of the FWI become inputs to the neural network, whereas the corresponding true 1D models are the output. The network is trained in a regression manner to detect the \textbf{BoS} and estimate the subsalt velocity. We analyze three scenarios in creating the training datasets and test their performance on the 2D BP 2004 salt model. We show that when the network succeeds in estimating the subsalt velocity, the requirement of low frequencies and long offsets are somewhat mitigated. In general, this work allows us to merge the top-to-bottom approach with FWI, save the \textbf{BoS} picking time, and empower FWI to converge in the absence of low frequencies and long offsets in the data.   
\end{abstract}

\keywords{ Full waveform inversion, deep learning, salt inversion, unflooding.}

\section*{Introduction}
Accurate imaging for salt bodies is of huge interest as subsalt sediments can hold potential hydrocarbon reservoirs. Salt bodies are mechanically light rocks that flow toward the surface when stress is applied and a conduit exists \citep{leveille2011subsalt}. They override the sediments forming large structures with steep boundaries and high acoustic impedance contrast. This complex nature of the salt bodies creates a challenge in imaging. Most of the energy of the propagated wave will reflect back from the salt's surface leading to insufficient illumination for the subsalt sediments. A successful subsalt imaging often requires integration of geological interpretation and advanced geophysical technology such as broadband acquisition, long offsets, high-quality multiple suppression, and advanced imaging techniques \citep{jones2014seismic,esser2016constrained}. 
\\
\\
Building a high-quality velocity model plays an important role in successful subsalt imaging. Full-waveform inversion (FWI) \citep{tarantola1984inversion,virieux2009overview} uses the full waveform information to obtain a high-resolution velocity model. It is implemented by minimizing the least square error between the observed and synthetic seismic data. The limited-band nature of the seismic data often leads FWI to be cycle skipped: a problem that occurs when the difference between the observed and the synthetic data is more than half a cycle, which leads FWI to fall into a local minimum. The basic approaches to address the cycle-skipping problem is to start with a good initial model \citep[e.g.,][]{datta2016estimatinginitial,ao2015sourceinitial,alali2020effectiveness} and to apply a multi-scale strategy \citep{bunks1995multiscale}. The multiscale strategy implies that the inversion is initiated with low frequency data and high frequencies are gradually introduced. The cycle-skipping problem can also be mitigated by applying preconditioning, regularization, or using more robust misfit functions \citep[e.g.][]{chi2014full,warner2016adaptive,kalita2019regularized,MF_OTMF}
\\
\\
In complex salt regions, the non-linearity issue is more severe. The absence of the salt bodies in the initial model will inevitably lead FWI to cycle-skipped solutions. The conventional industry approach to build a salt velocity model is the "top-to-bottom" workflow, which consists of multiple steps of imaging and manual picking for salt's boundaries \citep{dellinger2017garden}. The workflow starts by building the best sediment velocity using tomography and shallow diving wave FWI. Then, the top of the salt (\textbf{ToS}) is picked from the seismic image to flood the salt velocity in depth. After that, another imaging process is carried out to pick the bottom of the salt (\textbf{BoS}) and unflood the salt velocity at presumably the correct depth of the \textbf{BoS}. In complex models, the workflow often requires many iterations of flooding-unflooding, and testing different scenarios. The final salt model is determined by the quality of the final imaging. Although this approach often provides satisfying results, it is time-consuming and prone to errors as it highly depends on the interpreters' skills. This workflow sometimes is followed by FWI to correct for the erroneous picks of the salt body. 
\\
\\
Many successful salt inversions rely on data with low frequencies and long offsets. \cite{shen2017salt} implemented FWI using frequencies down to 1.6 Hz, full-azimuth, and long offset of about 20 km of ocean bottom nodes to invert for salt bodies in the Atlantis field. Their successful inversion reveals some misinterpretation of the legacy salt model, which was built using the "top-to-bottom" technique and used as an initial model for FWI. Similarly, \cite{wang2019saltComingAge} inverted for salt bodies in Keathly Canyon in the central GOM using frequencies down to 2.5 Hz, as a result, corrected for correct for the poor model obtained by the conventional approach. \cite{kumar2019updating} Compared two WAZ surveys in a salt inversion project and showed that the survey with the longer offset and lower frequency provided a more accurate salt body. \cite{kalita2019regularized} discussed the limitation of inverting for a salt model using vintage field data with limited offset and minimum usable frequency of 7 Hz. 
\\
\\
The recent rise in deep learning techniques unlocked a wide area of research in many geophysical applications, including seismic modelling, processing, inversion, and interpretation \citep[e.g.][]{waldeland2018interpretation,zeng2019automatic_salt,kazei2021mapping,song2021solving}. Convolutional neural networks (CNN), in particular, are robust in image processing and object detection problems, which attracted geophysicists to use them in interpretation tasks. In salt body inversion and reconstruction: \cite{lewis2017deep} extracted prior salt information from seismic images by CNNs, and use them in the velocity inversion, \cite{gramstad2018automatedTopBase} proposed to use two CNNs, one for picking the \textbf{ToS} and one to pick the \textbf{BoS} after sediment flooding, \cite{waldeland2017salt} and \cite{waldeland2018interpretation} used CNNs to classify the salts in seismic images. U-net, a specific CNN architecture that is mostly used for segmentation tasks, achieves high accuracy for salt interpretation \citep{zeng2019automatic_salt,shi2019saltseg,sen2020saltnet}.
Generally, most of the deep learning applications related to salt models are on the interpretation side, where the final image is already obtained and very little within the velocity building process. However, seismic images during the velocity building process are often corrupted, contaminated with noise, and thus, salt bodies are often poorly imaged. These factors make it difficult for CNN to interpret the salt accurately \citep{sen2020saltnet}.
\\
\\
The high scattering energy at the \textbf{ToS} and the lack of illumination at the \textbf{BoS} make the flooding step relatively easier than the unflooding step. Moreover, FWI can develop the \textbf{ToS} without any human intervention, which can be further utilized to apply an automatic flooding \citep{esser2016constrained,kalita2019regularized}. Therefore, here, we focus only on the unflooding problem and assume that the salt flooding is properly implemented. Our work focuses on two components: 1) unflooding the salt at the correct depth, 2) mitigating the requirement for having low frequencies and long offsets in doing so. The latter is achieved by obtaining an approximate of the subsalt velocity. We propose to apply FWI on the flooded model and then automate the unflooding using the U-net architecture in a regression framework. The network task is to unflood the salt and predict an approximation of the subsalt velocity. We show that if the network successfully approximates the subsalt velocity, then the requirement of having low frequencies and long offsets in FWI is reduced.  
\\
\\
The flow of the paper is as follows. We will start with an overview of FWI theory and the challenges in building a salt model. Then, we will demonstrate our method to unflood the salt and provide more details about the network and how we create the training datasets. After that, we will perform different experiments and analyze their performance on the left part of the BP 2004 salt model.
\section*{FWI}
Full-waveform inversion (FWI) is an optimization technique that aims to recover a high-resolution subsurface velocity \citep{tarantola1984inversion,virieux2009overview}. In its conventional form, it minimizes the squared $L_2$ norm of the difference between observed seismic data ($\mathbf{d_o}$), and the synthetic ones ($\mathbf{d_{syn}}$) obtained from the model $\mathbf{m}$. The objective function is given by:   
\begin{equation}
    J_{FWI}(\mathbf{m}) = \norm{\mathbf{d_o - d_{syn}(m)}}^2_2.
    \label{eq:fwi}
\end{equation}
Despite the success of FWI on many field data sets \citep{prieux2013multiparameter,plessix2010application,bansal2013simultaneousSourceFWI,choi2015unwrapped}, it still faces many challenges. FWI is based on linearized updates corresponding to the single scattering assumption, also known as the Born approximation; thus, model perturbations are limited. As a result, FWI fails to reconstruct the model using a poor initial model. The objective function $J_{FWI}$ is characterized as non-linear with multiple local minima, which create a serious challenge for gradient-based algorithms to resolve the global solution \citep{sirgue2004efficient_selectfreq,mulder2008exploring}. 
\\
\\
At low frequencies, $J_{FWI}$ tends to be more convex with less local minima \citep{ovcharenko2018variance}. A multiscale approach to invert for successive bands of frequencies starting from low to high is often applied to ensure a stable convergence \citep{bunks1995multiscale,brenders2007waveformpratt}. The low frequencies often develop the low-wavenumbers components of the model containing the large-scale variations. Carefully including the higher frequencies in the inversion will result in high-resolution details. Updating the low-wavenumbers can also be attributed to having a long offset \citep{vigh2013long,sirgue2006importancelong}. In many cases, seismic data lack the low frequencies and the wide offset needed to efficiently build the low-wavenumber components of the model, leading to unstable convergence of FWI, especially in complex models. 
\\
\\
To mitigate FWI limitations, many studies suggest using more robust objective functions that are more immune to cycle-skipping \citep[e.g.,][]{chi2014full,choi2015unwrapped,warner2016adaptive,alkhalifah2019efficient}. Others propose to impose some geological knowledge in the inversion as regularization or constraints \citep[e.g.,][]{asnaashari2013regularized,esser2016constrained,xue2017full}. One of the common regularizations used is total variation (\textit{\textbf{TV}}) which preserve the sharp edges of layers \citep{loris2012iterativeTV,esser2016constrained,kazei2017salt,kalita2019regularized}. The (\textit{\textbf{TV}}) regularization is written as: 
\begin{equation}
    \norm{\mathbf{m}}_{TV} = |\nabla \mathbf{m}| = \sqrt{\frac{\partial\mathbf{m}}{\partial x}+\frac{\partial\mathbf{m}} {{\partial z}}+\epsilon}.
    \label{eq:tv}
\end{equation}
Since (\textit{\textbf{TV}}) is non quadratic and has a singularity in its gradient, a small number $\epsilon$ is added to stabilize the problem.
\\
\\
In salt regions, building a velocity model using FWI is unfeasible in the absence of salt bodies in the initial model, low frequencies, and long offset \citep{ovcharenko2018variance}. At best, FWI will construct the post-salt sediments and the \textbf{ToS} boundary, but often yields poor Subsalt information, including the BoS. Therefore, the traditional practice aims to include the salt boundaries (\textbf{ToS} and \textbf{BoS}) in the initial model by means of the top-to-bottom workflow, then follow that with low frequencies and long-offsets FWI inversion. A more advanced approach is based on the fact that FWI can capture the \textbf{ToS}, which can allow for an automatic flooding \citep{esser2016constrained,kalita2019regularized,alali_salt_ml}. However, these approaches cannot detect the \textbf{BoS} if frequencies are not low enough and offsets are not long enough, and, thus, still requires human intervention to control the amount of flooding.

\section*{Method}
\begin{figure}[!h]
    \centering
    \includegraphics[width=.9\textwidth]{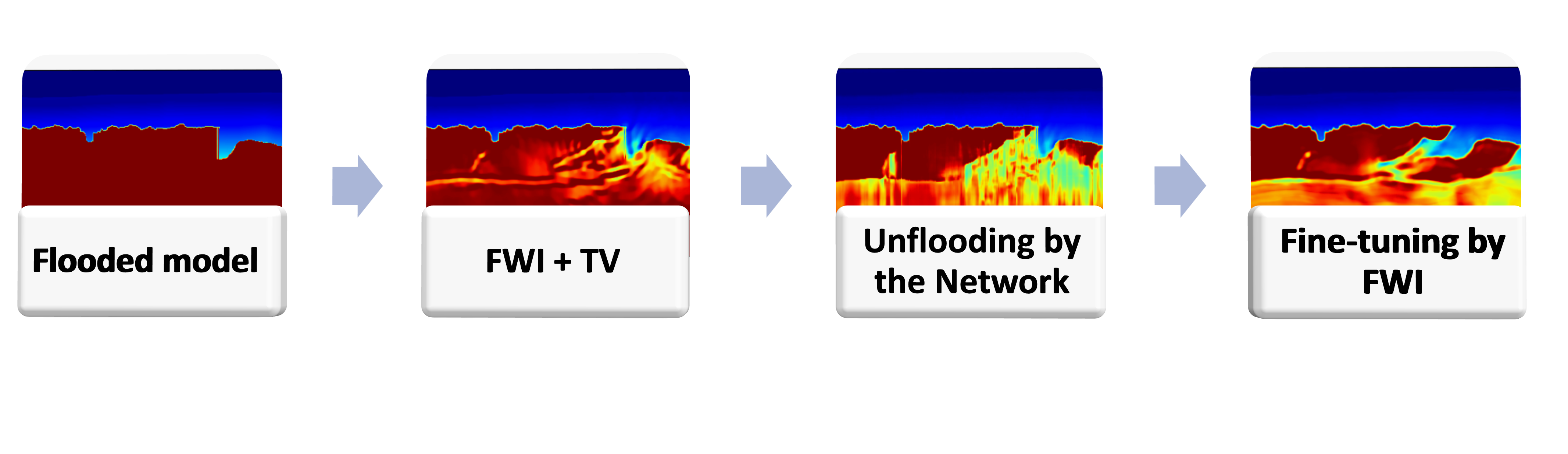}
    \caption{Workflow of the proposed method. Starting with a flooded model, we apply FWI with TV. Then, the network will unflood the model. Finally, another FWI is implimented to fine-tune the model.}
    \label{fig:workflow}
\end{figure}

Here, we provide an approach to invert for the subsalt velocity in the challenging case of missing low frequencies and limited offsets. We assume that the post-salt sediments are well-recovered and the \textbf{ToS} is picked correctly for flooding; thus, our main focus is to detect the \textbf{BoS} and provide a good prediction of the subsalt sediments.
\\
\\
In salt inversion, FWI is often combined with \textit{\textbf{TV}} regularization to preserve the sharp edges of the salt boundaries. Applying FWI (equation~\ref{eq:fwi}) with \textit{\textbf{TV}} (equation~\ref{eq:tv}) to the flooded model ($\mathbf{m_{flooded}}$) yields:  
\begin{equation}
        J_{FWI+TV}(\mathbf{m_{Flooded}}) = J_{FWI}(\mathbf{m_{Flooded}})  + \lambda \norm{\mathbf{m_{Flooded}}}_{TV},
        \label{eq:fwitv}
\end{equation}
here $\lambda$ is a coefficient balancing the contribution of the two norms. Minimizing the objective $J_{FWI+TV}(\mathbf{m_{flooded}})$ should reveal some features of the \textbf{BoS}, such as a sharp drop in the velocity at the \textbf{BoS} depth. These features (velocity drops) can be automatically captured using a neural network to unflood the salt at the \textbf{BoS}. If the network is trained in a regression fashion, the network will estimate the subsalt velocity allowing a better inversion in a subsequence FWI. Having a good initial subsalt velocity reduces the requirement of using long offsets and low frequencies. We illustrate this workflow in Figure~\ref{fig:workflow}.
\subsection*{The neural network setup}
\begin{figure}[!h]
    \centering
    \includegraphics[width=.6\textwidth]{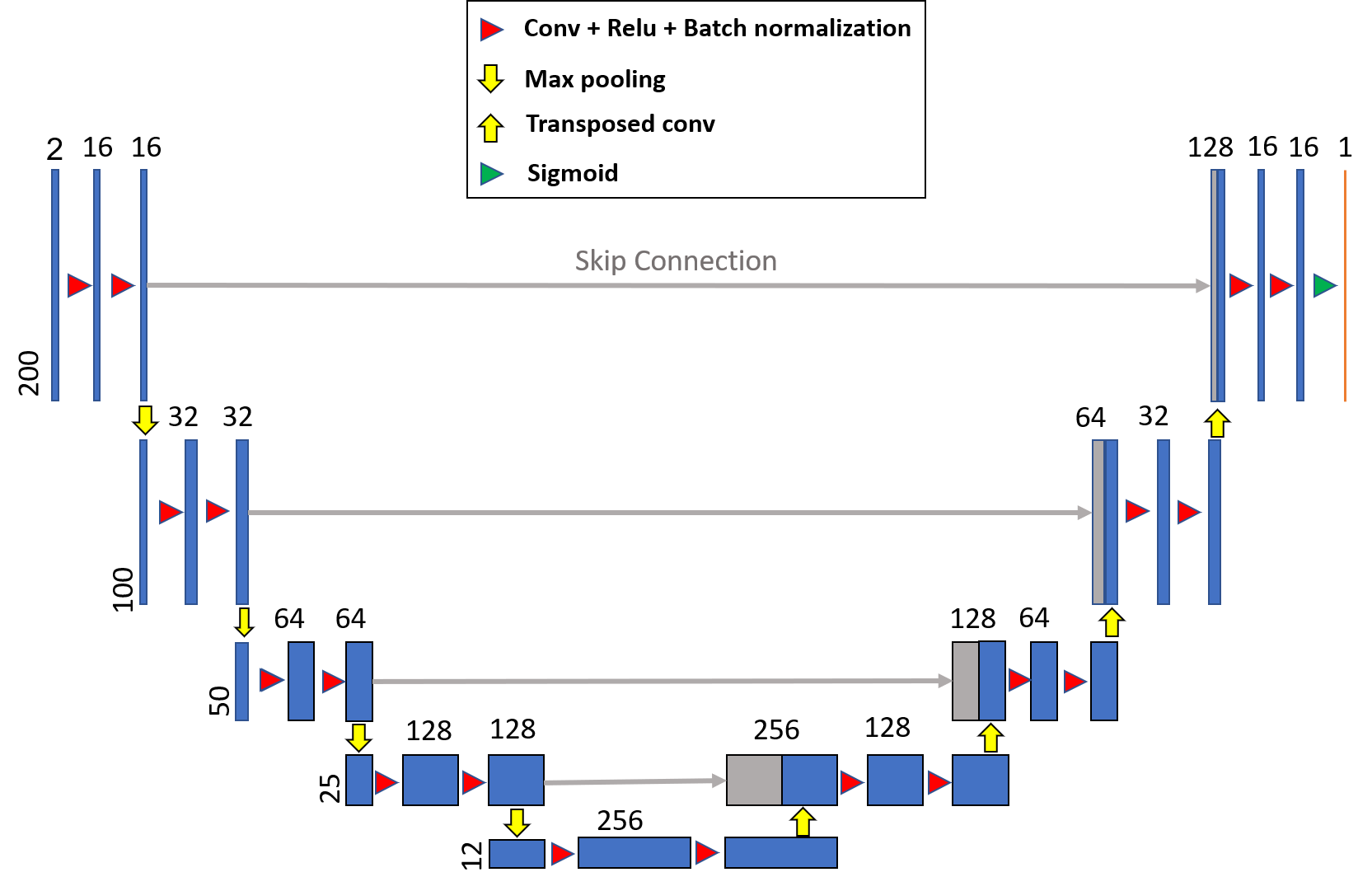}
    \caption{1D U-net architecture.}
    \label{fig:unet}
\end{figure}

To unflood the salt, we adapt a U-net architecture as our neural network. U-net is commonly used in segmentation tasks and has been widely used for salt detection as a classifier. It consists of symmetric blocks of encoders, that compress the size of the input while extracting features, and decoders to recover the size. Normally U-net is applied on 2D images. Here, we modify the structure to accept 1D inputs by using 1D convolutional layers instead of 2D. This network is summarized in Figure~\ref{fig:unet}. It is composed of four encoder and four decoder blocks. In each block, we have two convolutional layers with a rectified linear unit (Relu) activation function followed by a batch normalization operator. The encoder blocks are connected to each other by a max-pooling operator to compress the size. In the decoders, transposed convolutional layers are used to recover the size of the input. Typical U-net has an increasing number of channels in the encoder and a decreasing number in the decoder by a factor of 2. The number of channels used in this work are 16, 32, 64, 128 and 256 in the bottleneck. Finally, a sigmoid activation function is applied on the last layer to produce the output. 
\\
\\
The input to the network consists of two channels: the FWI result of the flooded model (i.e., the solution of equation~\ref{eq:fwitv}), and the initial model used for the inversion. Including the initial model in the input affects the prediction in areas that do not contain salts by using information from the initial model. The output of the network will be the unflooded model. The unflooding process can be expressed by: 
\begin{equation}
    \mathbf{m_{Unflooded}} = \theta(\mathbf{m_{FWI}}),
    \label{eq:predict}
\end{equation}
where $\mathbf{m_{FWI}}$ is the FWI inversion for the flooded model, and $\theta$ indicates the network. For the training, we use the true model $\mathbf{m_{True}}$ as our target. The network is trained by minimizing the mean-squared-error (MSE) loss: 
\begin{equation}
    \theta_{loss} = \frac{1}{N} \sum^N_{i=1} (\mathbf{m_{Unflooded}}-\mathbf{m_{True}})^2, 
    \label{eq:loss}
\end{equation}
where $N$ denotes the batch size. Unlike the commonly used cross-entropy loss for classification that will only be useful to capture the salt boundaries, MSE will also approximate the subsalt velocity. 

\subsection*{Training dataset}
Training a network requires large and diverse datasets to cover all the possible scenarios in the application. The input to our network is FWI results; thus, we will need many FWI implementations on various plausible models. Therefore, we focus the training on one dimensional models, where the FWI requires modeling a single shot gather. This should be fine considering that flooding and unflooding are often handled along the vertical dimension. Also, in 1D models, the gradient can be computed using a few shots by laterally stacking the gradient, which simulates the effect of a full fold coverage \citep{alkhalifah2018singleshot}. We create 8000 random layered models with a minimum of five layers. To mimic a marine setup, the models start with the water velocity of 1.5 km/s and generally increase with depth. We randomly smooth several profiles to increase the diversity of the training samples. Then, a salt layer is randomly added to a number of the samples with a velocity of 4.5 km/s. The salt thickness and position are also randomly determined. Leaving some of the samples without salt bodies teaches the network to differentiate profiles containing salts from those that do not. 
\\
\\
To apply FWI with TV on the generated models, we synthesize one shot for each model using a Ricker wavelet of 8 Hz dominant frequency. The target seismic data described in the next section has a maximum offset up to 5 km and a minimum frequency of 5 Hz, which is often not suitable to correct for the flooded subsalt. So we train the neural network with similar dimensions. Since we assume that the post-salt model is well known, we use the true models except beneath the salt, where we flood the salt velocity, as our initial model for FWI. The inversion results are then used to train the network as we will show in the following section.


\section*{Experiments}
\begin{figure*}[!h]
    \centering
    \includegraphics[width=0.6\textwidth]{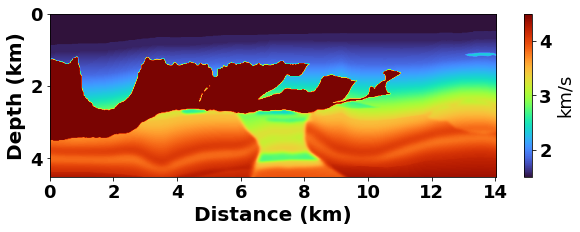}
    \caption{The left part of the BP 2004 salt model.}
    \label{fig:BP_t}
\end{figure*}

\begin{figure*}[!h]
    \centering
    \subfigure[]{\includegraphics[width=0.4\textwidth]{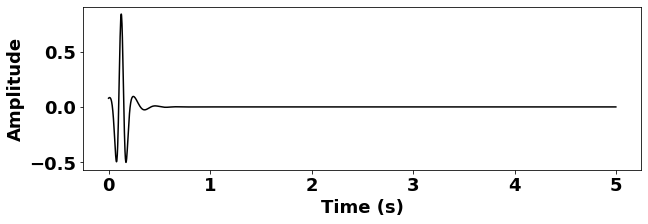}
    \label{fig:wavelet}}
    \subfigure[]{\includegraphics[width=0.4\textwidth]{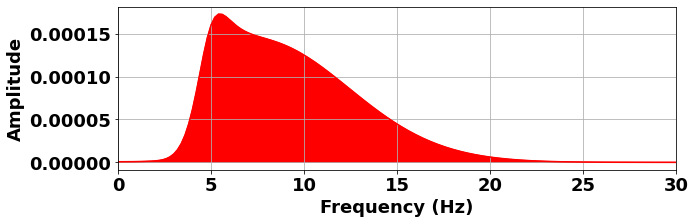}
    \label{fig:freq_wavelet}} 
    \caption{(a) The Ricker wavelet used to simulate the data for the BP model and (b) its frequency content. Note the low frequencies are missing.}
    \label{fig:Allwavelet-freq}
\end{figure*}

\begin{figure*}[!h]
    \centering
    \subfigure[]{\includegraphics[scale=0.3]{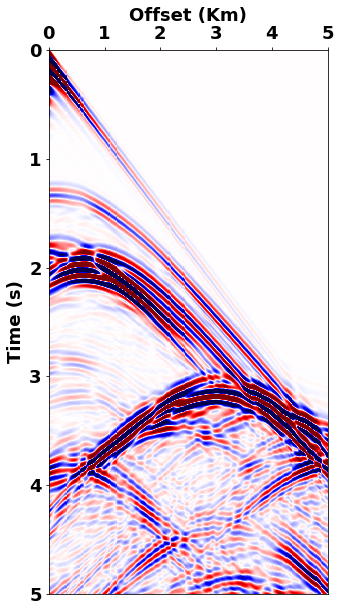}
    \label{fig:shot}}
    \subfigure[]{\includegraphics[width=0.4\textwidth]{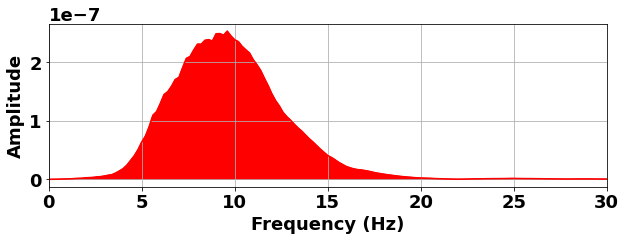}
    \label{fig:freq_shot}} 
    \caption{(a) A shot gather from the source at position 0 km and (b) its average spectrum. We can see that the data are missing the low frequencies.}
    \label{fig:Allshot-freq}
\end{figure*}

\begin{figure*}[!h]
	\centering
	\subfigure[]{\includegraphics[width=0.6\textwidth]{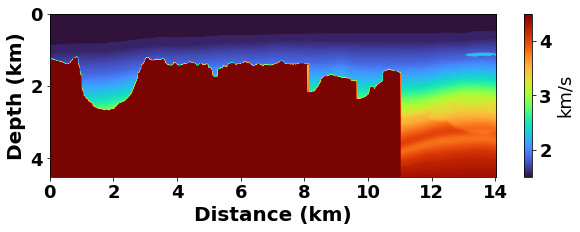}
		\label{fig:BP_f}} \vspace*{-0.03\columnwidth} \vspace*{0pt}
	\subfigure[]{\includegraphics[width=0.6\textwidth]{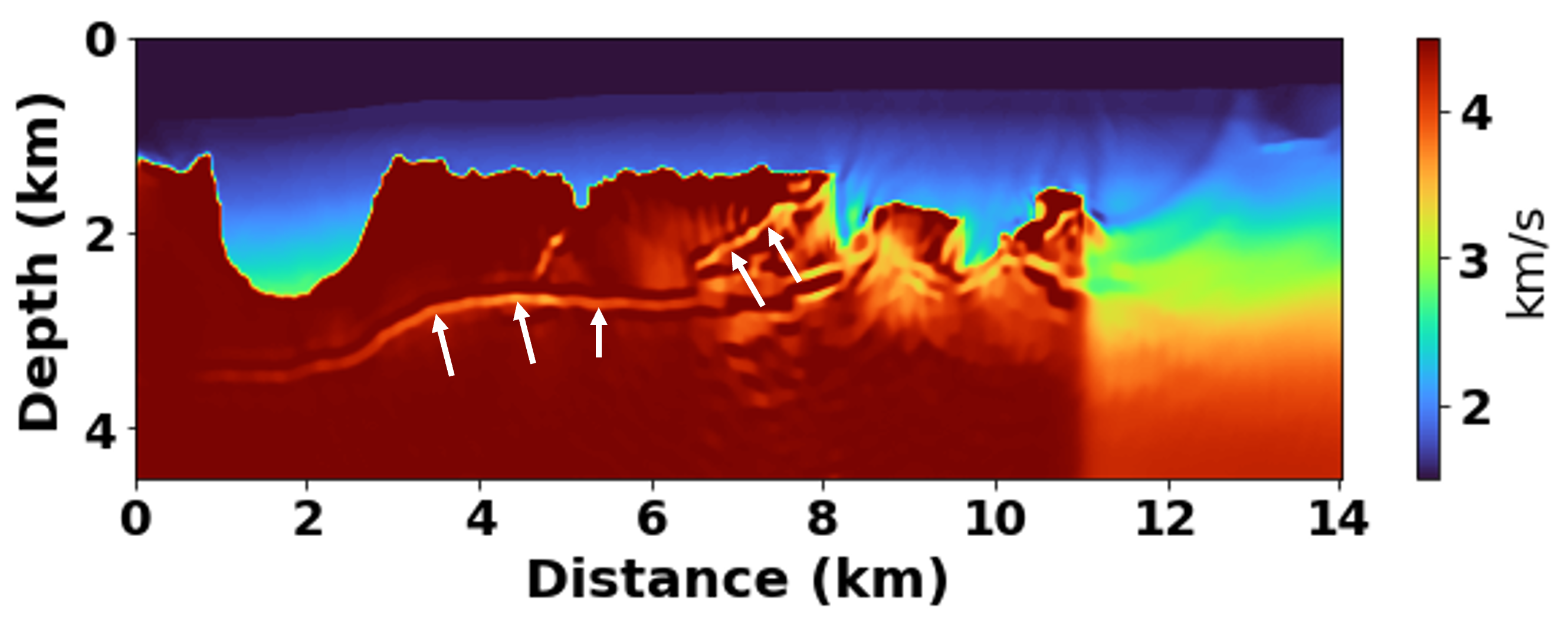}
		\label{fig:BP_inv}} \vspace*{-0.03\columnwidth}
	\caption{(a) The flooded BP model that is used as an initial model in FWI, and (b) is the result of the inversion. FWI did not manage to invert for the subsalt velocity due to the lack of low frequencies and using short offset. However, we can observe that the \textbf{BoS} can be easily identified now as indicated by the white arrows.}
	\label{fig:BP_initial-fwi}
\end{figure*}

\begin{figure*}[!h]
    \centering
    \subfigure[1st experiment's loss]{\includegraphics[width=0.3\textwidth]{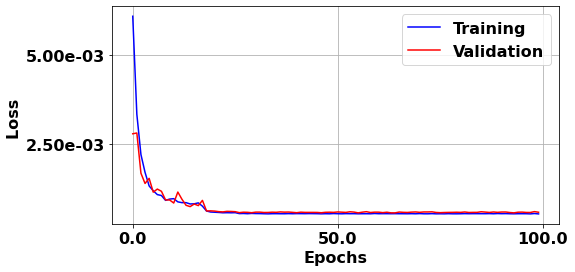}
    \label{fig:loss1}} 
    \subfigure[2nd experiment's loss]{\includegraphics[width=0.3\textwidth]{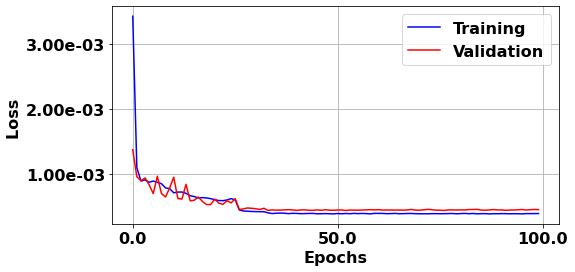}
    \label{fig:loss2}} 
    \subfigure[3rd experiment's loss]{\includegraphics[width=0.3\textwidth]{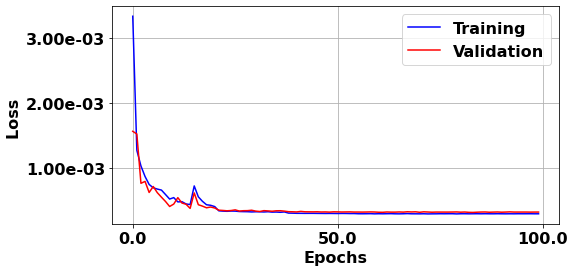}
    \label{fig:loss3}}     
    \subfigure[1st experiment's R2 score]{\includegraphics[width=0.3\textwidth]{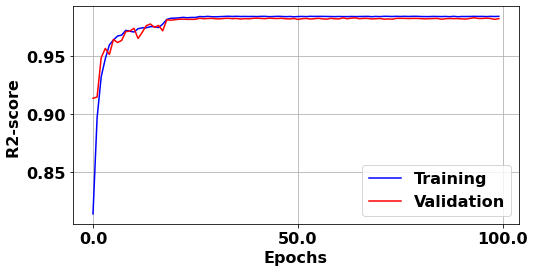}
    \label{fig:R2-1}} 
    \subfigure[2nd experiment's R2 score]{\includegraphics[width=0.3\textwidth]{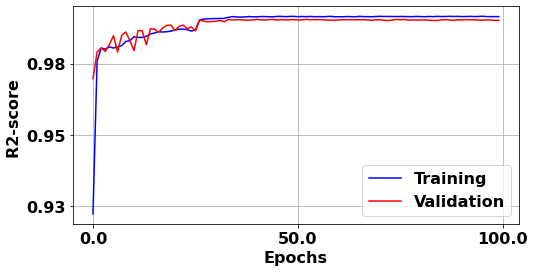}
    \label{fig:R2-2}} 
    \subfigure[3rd experiment's R2 score]{\includegraphics[width=0.3\textwidth]{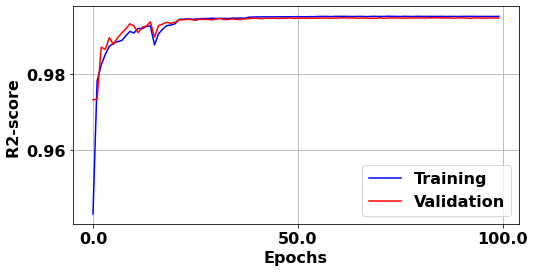}
    \label{fig:R2-3}}     
    \caption{Losses and R2 scores for the three experiment. Each column shows the loss and R2 score for one of the experiments, starting from the first experiment to the third one.}
    \label{fig:losses-R2}
\end{figure*}

\begin{figure*}[!h]
    \centering
    \subfigure[]{\includegraphics[width=0.9\textwidth]{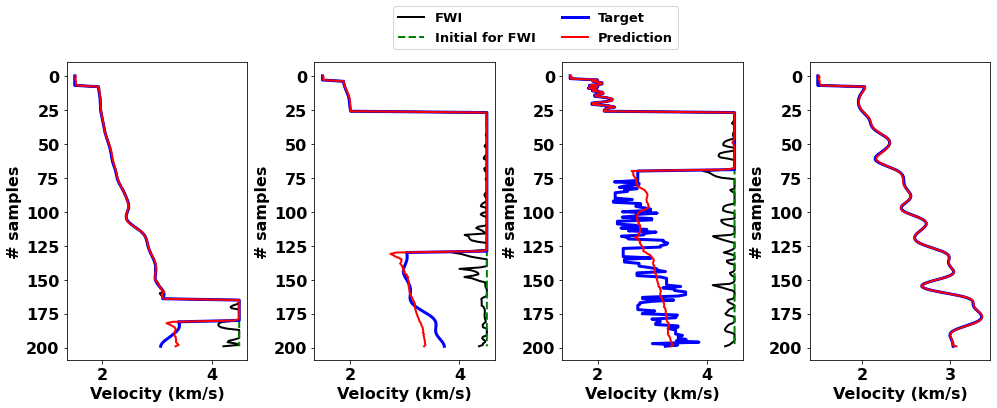}
    \label{fig:samples_1}} 
    \subfigure[]{\includegraphics[width=0.9\textwidth]{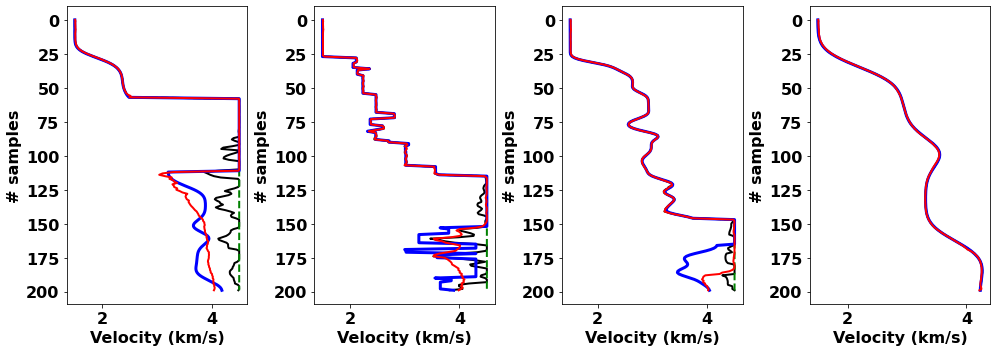}
    \label{fig:samples_2}} 
    \subfigure[]{\includegraphics[width=0.9\textwidth]{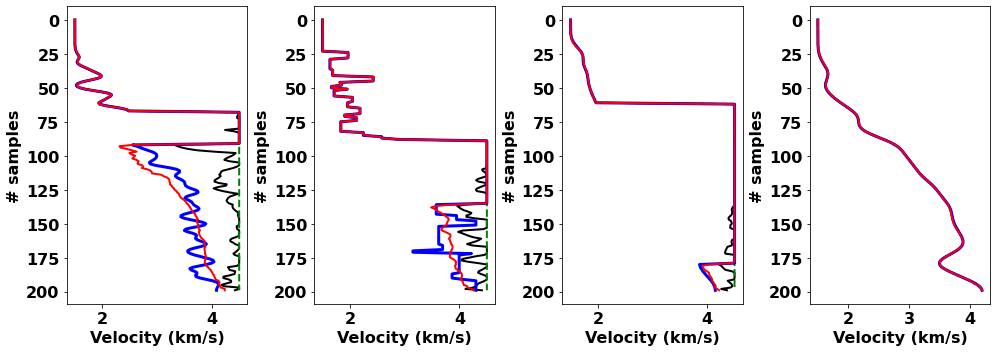}
    \label{fig:samples_3}}     
    \caption{Samples from the validation set and their corresponding unflooding by U-net for (a) the first experiment, (b) the second experiment and (c) the third experiment. The black and the green dashed lines are the FWI results and their initial models used as inputs to the network. The red line is the unflooding prediction by the U-net and the blue line is the target given by the true model. When the model does not contain a salt such as the samples in the last column, FWI and its initial model are the same as the target.}
    \label{fig:samples_all}
\end{figure*}

\begin{figure*}[!h]
	\centering
	\subfigure[]{\includegraphics[width=0.6\textwidth]{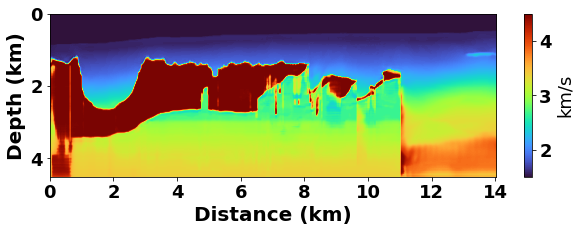}
		\label{fig:BP_unf1}} \vspace*{-0.03\columnwidth} \vspace*{0pt}
	\subfigure[]{\includegraphics[width=0.6\textwidth]{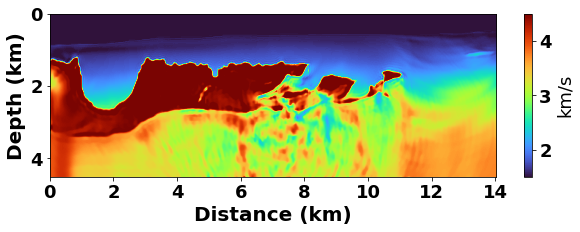}
		\label{fig:BP_inv2_1}} \vspace*{-0.03\columnwidth}
	\caption{(a) The unflooding for the BP model using the network of the first experiment and (b) a follow up FWI to fine-tune the model. The network captures the \textbf{BoS} but it fails to predict the subsalt velocity. As a result, the fine-tuning FWI also fails to build the subsalt velocity despite the improvements in recovering the salt body.}
	\label{fig:BP_1stExp}
\end{figure*}

\begin{figure*}[!h]
	\centering
	\subfigure[]{\includegraphics[width=0.6\textwidth]{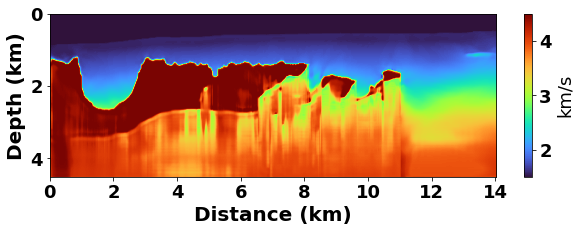}
		\label{fig:BP_unf2}} \vspace*{-0.03\columnwidth} \vspace*{0pt}
	\subfigure[]{\includegraphics[width=0.6\textwidth]{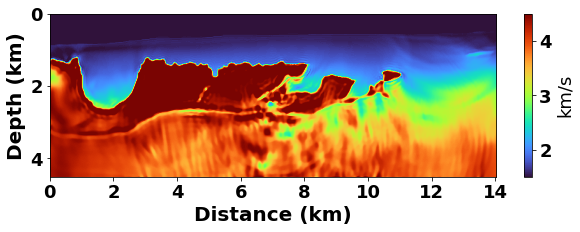}
		\label{fig:BP_inv2_2}} \vspace*{-0.03\columnwidth}
	\caption{(a) The unflooding for the BP model using the network of the second experiment and (b) a follow up FWI to fine-tune the model. The network captures the \textbf{BoS} and also approximates the subsalt velocity, which makes the FWI reconstructs the subsalt model better even with the limited offsets and the lack of low frequencies.}
	\label{fig:BP_2ndtExp}
\end{figure*}

\begin{figure*}[!h]
	\centering
	\subfigure[]{\includegraphics[width=0.6\textwidth]{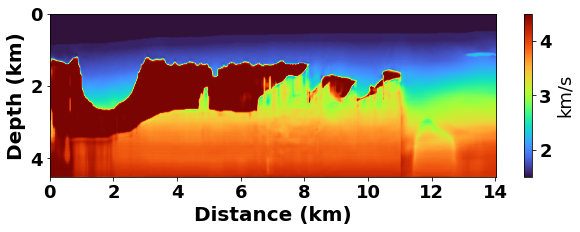}
		\label{fig:BP_unf3}} \vspace*{-0.03\columnwidth} \vspace*{0pt}
	\subfigure[]{\includegraphics[width=0.6\textwidth]{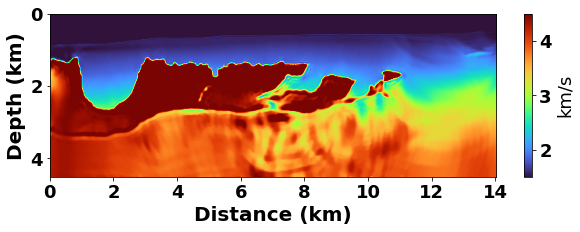}
		\label{fig:BP_inv2_3}} \vspace*{-0.03\columnwidth}
	\caption{(a) The unflooding for the BP model using the network of the third experiment and (b) a follow up FWI to fine-tune the model. Similar to the second experiment, the network manages to estimate the subsalt velocity enabling the subsequent FWI to recover the subsalt velocity despite the limitation of the data.}
	\label{fig:BP_3rdExp}
\end{figure*}

\begin{figure*}[!h]
    \centering
    \subfigure[]{\includegraphics[width=0.2\textwidth]{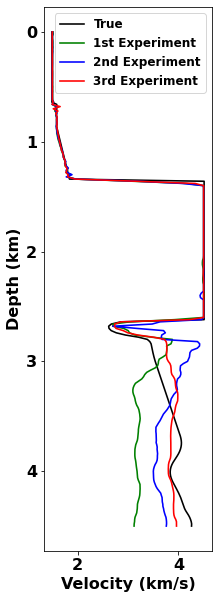}}
    \subfigure[]{\includegraphics[width=0.2\textwidth]{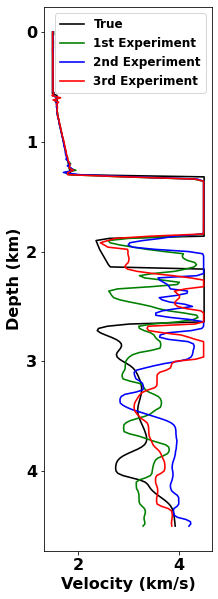}}
    \caption{Vertical velocity profiles taken at (a) 4.5 km and (b) 7.3 km (right). The black line indicates the true model, while the green, blue and red lines indicate the final inversion of the first, second and third experiment, respectively. In most cases, the inversion in the subsalt area captures the low velocity layer in (a), and shows hints for the low velocity region in (b), especially for the third experiment.}
    \label{fig:prof}
\end{figure*}

We test the proposed method on the left part of the BP 2004 benchmark model, shown in Figure~\ref{fig:BP_t}. We down-sample the BP model to half its original size while keeping the original aspect ratio. We use the constant-density wave equation to simulate 200 shots on the surface separated by 70 m. The data were recorded, as mentioned earlier, with a limited offset of 5 km. A Ricker wavelet with 8 Hz dominant frequency is used as a source (Figure~\ref{fig:wavelet}). A high-pass Butterworth filter is applied with a cutoff frequency of 5 Hz to remove low frequencies. The frequency spectrum for the wavelet is plotted in Figure~\ref{fig:freq_wavelet}. The maximum offset of the streamer is 5 km with 20 m receiver spacing. Figures~\ref{fig:shot} and \ref{fig:freq_shot} show a shot gather obtained from a source positioned at 0 km and its average frequency spectrum, respectively. The spectrum further indicates that the data are missing the low frequencies. We then perform FWI with TV on a flooded version of the model shown in Figure~\ref{fig:BP_f}. This inversion is implemented using all available frequencies of the data without the multiscale approach as the purpose is to obtain an imprint of the \textbf{BoS}. Figure~\ref{fig:BP_inv} shows the inversion result. As expected, FWI could not correct the subsalt velocity. However, we can see signatures of the \textbf{BoS} as indicated by the white arrows.    
\\
We ran three experiments. The difference between the three experiments lay in the training datasets. The general properties in generating the training dataset were provided in the previous section. Altering the mean and range of the velocity models can have a large impact on performance. Table~\ref{table:1} states the main differences between the three experiments. In the first experiment, the data were created in quasi-random models described above. In the second experiment, we include some features of the BP model in the training datasets such as, the water depth, the maximum sediment velocity and faster increase of the velocity with depth. To create the training data in the third experiment, we use the average depth values of the BP velocity, excluding the salt, as a background.     
\rowcolors{2}{gray!25}{white}
\begin{table}[h!]
\begin{center}
\begin{tabular}{ |p{3cm}||p{8cm}| } 
 \rowcolor{gray!25}
 \hline
 \textbf{Experiment 1} & The data is created arbitrary with no geological knowledge about the targeted BP model  \\ 
  \hline
 \textbf{Experiment 2} & Some features are modified to better match the BP model such as the water depth and the maximum velocity of the sediments \\ 
 \hline
 \textbf{Experiment 3} & The average velocity of the sediments, excluding the salt, were used as a general trend in creating the training datasets  \\
 \hline
\end{tabular}
\caption{Differences between the three experiments.}
\label{table:1}
\end{center}
\end{table}
\\
\\
The training hyper-parameters for all the experiments are the same. The data are divided such that 80\% are used for training and 20\% for validation. We start the training using 0.1 learning rate that decays when the validation reaches a plateau. We run the training for 100 epochs using a batch size of 32 samples. We observe similar convergence behaviors for training all the experiments, as we can see in the loss curves shown in the first row of Figure~\ref{fig:losses-R2}. We use the coefficient of determination, also known as the R2 score,
\begin{equation}
     R^2 = 1 - \frac{\sum_i^N (m_{true}-m_{unflood})^2}{\sum_i^N (m_{true}+\bar{m}_{true})^2},
\end{equation}
as an accuracy metric and achieve a score larger than 98\% in all the experiments. $\bar{m}_{true}$ denote the mean value of $m_{true}$. Figure~\ref{fig:samples_1}, \ref{fig:samples_2} and \ref{fig:samples_3} show some samples from the validation set for the first, second and third experiment, respectively. The green dashed lines are the initial models used in FWI. Since the initial models are the same as the true above the salt, we only see the green dashed lines in the flooding region. The resulting FWI plotted in black lines clearly indicate some signature for the \textbf{BoS}. We observe that FWI affects even the salt block in some samples. The unflooding predictions by the network (the red lines) fix the salts ruined by FWI, and capture the \textbf{BoS}. Comparing the predictions with the true models (the blue lines), we can see that the network predicts the general trend of the subsalt velocity. For the samples that do not contain salt (last column in Figure~\ref{fig:samples_all}), we input the true model in the network for both the FWI and its initial, which teaches the network to handle non-salt models. The accurate prediction of the post-salt sediments, salt layer and the non-salt models, which cover the majority of the training points, justify the relatively high R2 score we obtain.       
\\
We apply the three trained networks on the inverted BP model in Figure~\ref{fig:BP_initial-fwi}. The unflooding result by the first experiment is shown in Figure~\ref{fig:BP_unf1}. We see that the network captures the \textbf{BoS} in most regions, but it fails in estimating the subsalt velocity. Following the unflooding model with a final FWI improves the model but still cannot recover the subsalt velocity due to the missing low frequencies and the relatively short offsets. This final FWI is implemented with a multiscale approach by initially inverting for frequencies up to 7.5, 8.5, 10 and then the full bandwidth. Figure~\ref{fig:BP_unf2} is the unflooding result for the second experiment. We notice a dramatic improvement in the prediction for both the \textbf{BoS} detection and the subsalt velocity. Due to having 1D predictions, we observe vertical stripes in the prediction, especially in the subsalt area. The improvement in the unflooding helps the subsequent FWI to recover some of the subsalt model, as shown in Figure~\ref{fig:BP_inv2_2}. For the last experiment, we plot the unflooding and its fine-tuning FWI in Figures~\ref{fig:BP_unf3} and \ref{fig:BP_inv2_3}. Here, we do not observe the vertical stripes as in the second experiment. Instead, the unflooding is homogeneous below the salt with a velocity of about 4 km/s, which is about the average velocity of BP at that depth. The inversion recovers reasonable features of the subsalt model similar to the second experiment. Among these features, are hints of the low velocity region between locations 6 and 8 km, below the salt, as well as, a low velocity layer around 4 km. This can be observed in Figure~\ref{fig:prof} showing vertical velocity profiles at 4.5 and 7.3 km, in which we compare the results from the three experiments with the true model. The RMS differences ($\sqrt{\frac{1}{N}\sum_i^N (a_i-b_i)^2}$) between the true model and the final inversion in the third experiment is 0.26, which is lower than the second and the first experiments, that have RMS differences of 0.32 and 0.43, respectively. The R2 scores between the true model and the final inversions are 0.85 for the first experiment, 0.91 for the second and 0.94 for the third. The RMS and the R2 scores suggest that the final experiment achieve the best inversion among the three.   

\section*{Discussion}
\begin{figure*}[!h]
	\centering
	\subfigure[First Experiment]{\includegraphics[width=0.4\textwidth]{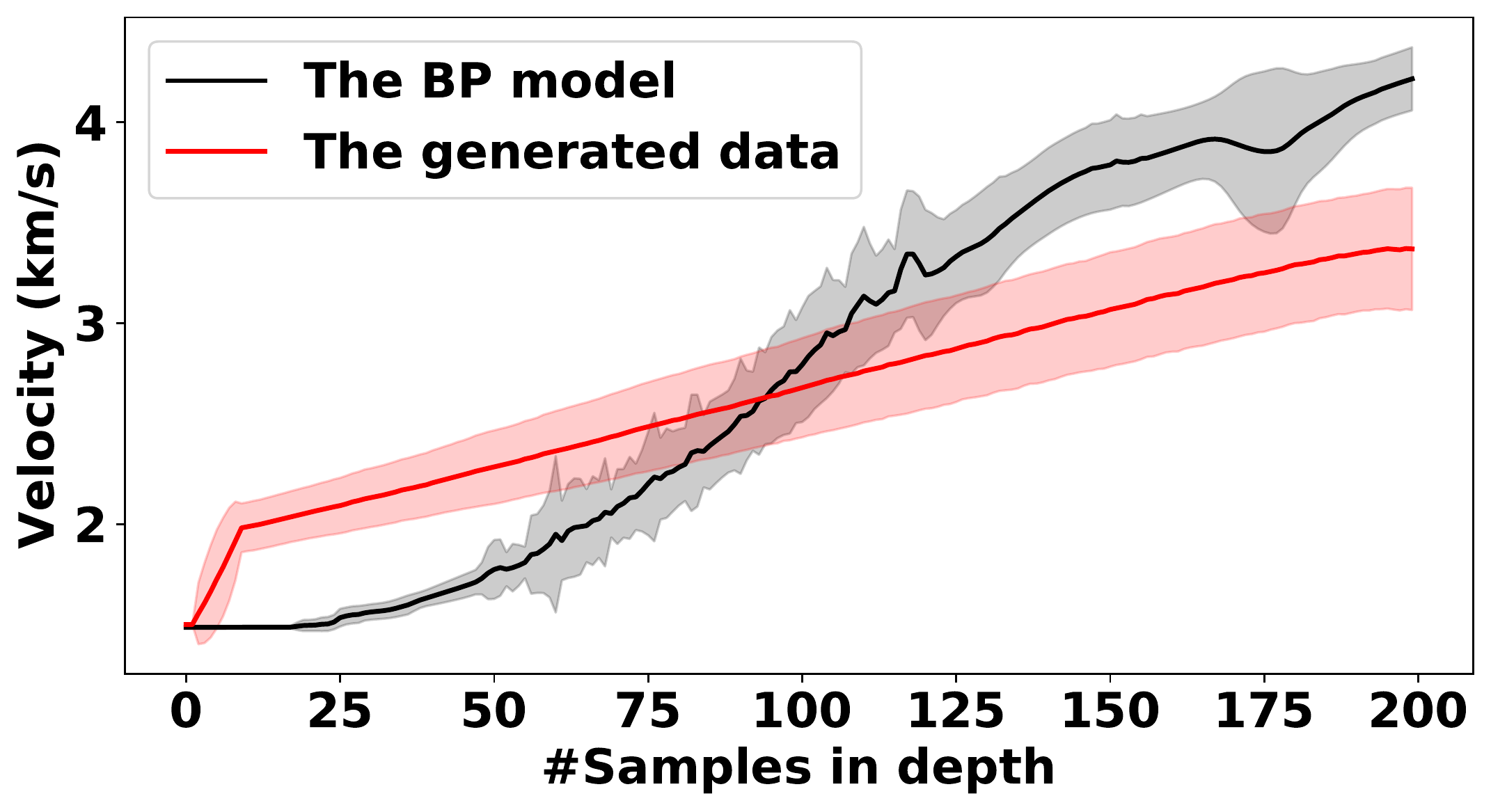}
		\label{fig:BPvsDATA_1}} 
	\subfigure[Second Experiment]{\includegraphics[width=0.4\textwidth]{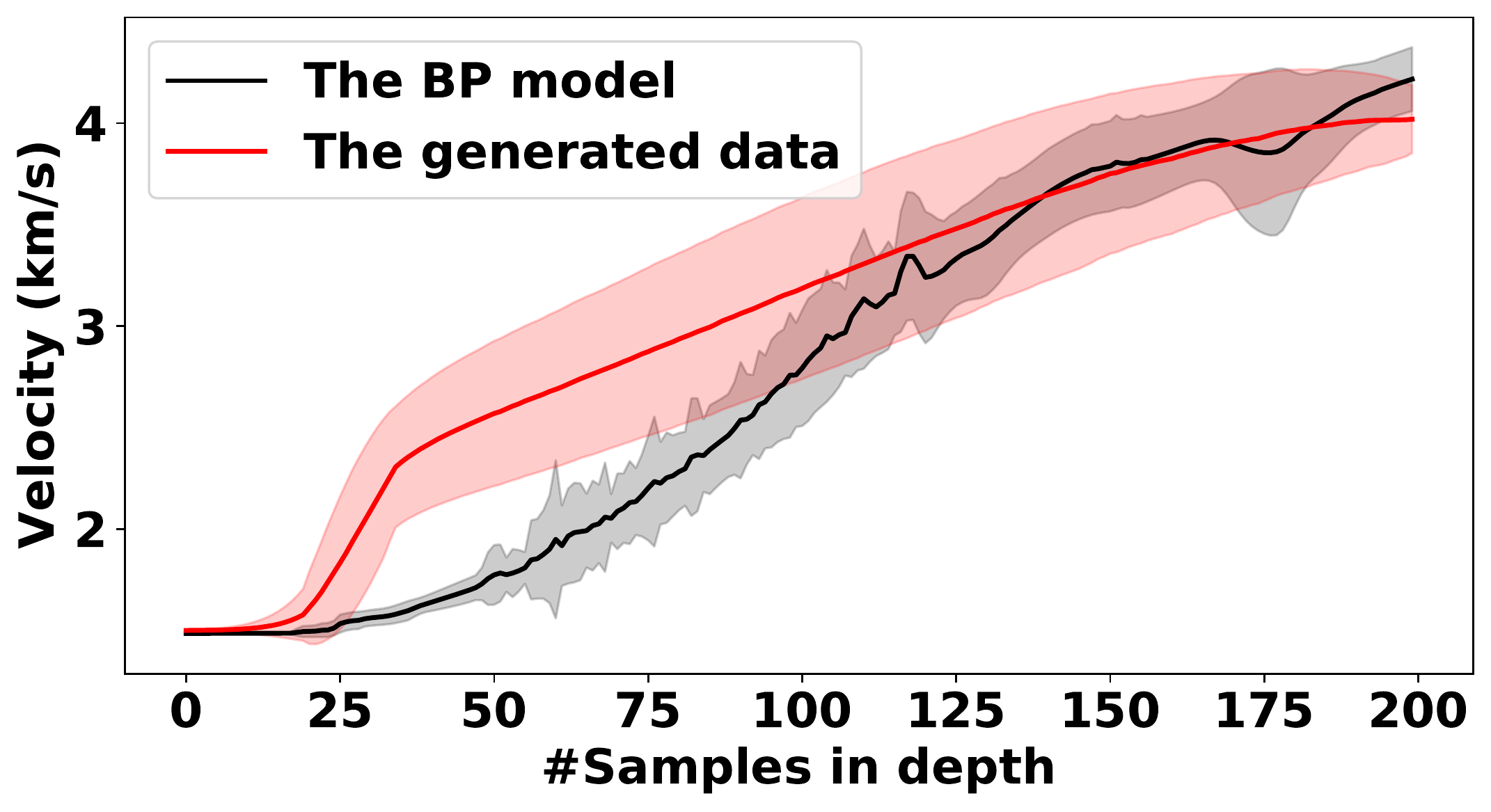}
		\label{fig:BPvsDATA_2}}
	\subfigure[Third Experiment]{\includegraphics[width=0.4\textwidth]{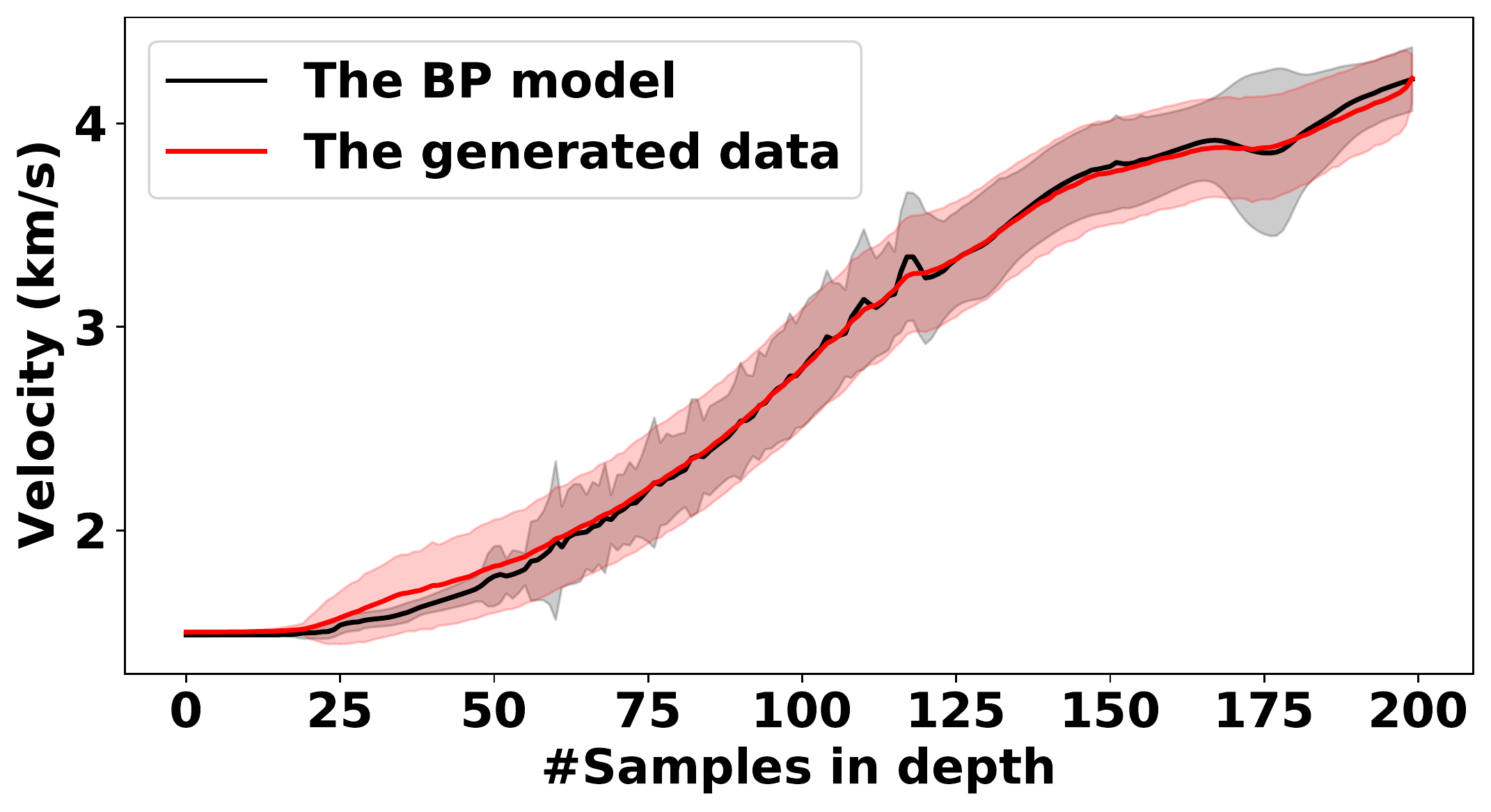}
		\label{fig:BPvsDATA_3}} 
	\caption{Comparison between the training data's means$\pm$standard-deviations (red) over depth and the BP model's means$\pm$standard-deviations (black) excluding the salt layers. The gap between the two lines in the first experiment (a) explains the bad prediction at the subsalt. In the second experiment (b) the training data cover the BP model curve in the subsalt region which justify the improvments in the subsalt prediction. In the last experiment (c), the training data and the BP models match each other.}
	\label{fig:BPvsDATA_all}
\end{figure*}

The performance of the NN for all three experiments, with the three datasets, is similar. However, they yield different results when applying the network on the BP model. This is due to the difference in the statistics between the generated data and the BP model. A simple test to compare the means$\pm$standard-deviations is shown in Figure~\ref{fig:BPvsDATA_all}. Note that these means$\pm$standard-deviations curves are excluding the salt layers. Ideally, the training dataset should cover all possible values in the test dataset, in our case the BP model. This implies that the red region in Figure~\ref{fig:BPvsDATA_all}, which represents the generated dataset, should contain the black region corresponding to the BP model. We see that this is not the case in the first experiment (Figure~\ref{fig:BPvsDATA_1}). In fact, the maximum velocity in the generated dataset is about 3.5 km/s, which is below the maximum sediment velocity of the BP model, possibly explaining why the network under-estimated the subsalt velocity in Figure~\ref{fig:BP_unf1}. In the second experiment (Figure~\ref{fig:BPvsDATA_2}), the generated dataset curve slightly deviates from the BP model after the water layer, but it starts to match it in the deeper half. This match between the training data and the BP model at depth justifies the improvements in the inverted BP model as this area is where the salt and the subsalt exist. The third experiment is constructed intentionally to match the means$\pm$standard-deviations for BP as shown in Figure~\ref{fig:BPvsDATA_3}. It presents a way to preserve the average background velocity of the BP model. In our implementation, we design the training dataset based on a particular model statistics, but we believe the network can be generalized better if more than one model statistics is used in generating the training datasets. Nevertheless, the 1D training approach we used is not expensive, and we promote gearing the training to handle a particular region and the corresponding seismic data. The statistics needed to generate the training set, in this case, can be deduced from any tomographic or FWI implementations in areas of the region free of salt. Our approach assumes that some processing and imaging have already been applied to the data, some of which was need to arrive to the flooded salt model.
\\
\\
In real-life scenarios, we suggest generating the training dataset using some prior knowledge of the area. Using an average velocity from some wells or a tomographic velocity as a general trend in generating the training dataset replicates experiment 3, which yields a homogeneous subsalt velocity. However, when the wells velocities are different from the subsalt, using partial information such as the expected maximum velocity is a safer choice, which is similar to experiment 2. However, in this case some horizontal smoothing might be necessary to remove the vertical stripes resulting from the network. 
\\
\\

The key factors for the success of our approach are to have the \textbf{ToS} accurately mapped and to apply an inversion that preserves the \textbf{ToS}. For the latter, we propose to apply TV regularization. Note the use of TV regularization is merely for edge preserving purposes and does not play a major role in the unflooding workflow. However, if it is not used in the inversion, it should not be used in the training data. Even with TV regularization, we see that at position =9 km in the first BP inversion (Figure~\ref{fig:BP_inv}), the salt boundaries were deformed by FWI. This affects the unflooding in all the experiments, leading the network to detect a false \textbf{BoS}. Another false unflooding by all the experiments is observed in the area between 7 and 8 km. This false unflooding is due to the fact that BP has two salt layers in that region and our network is only trained on dataset containing one. A solution to address the two-salt (or more) problem would be to include models with two salt layers (or more) in the training datasets. Following the predicted unflooded models with an FWI, assists in resolving some of these false predictions by the network. 
\\
\\
In the non-salt region between 11 and 14 km, the network only slightly modifies the model. This is because of the extra channel in the input (i.e., the initial model for the first inversion). The network is trying to balance the two channels. Not including the extra channel will keep the non-salt model unchanged in the prediction. Thus, we suggest adding the channel only when we trust our initial model to be representative of the velocity in the region. In our case the initial model is the true model without the salt body.
\\
\\
Unlike the conventional approach, where the \textbf{BoS} is picked from images, our approach detects it from the velocity inversion. While this seems to add more cost to the problem, it automates the detection of the \textbf{BoS}. More importantly, it provides a good approximation for the subsalt velocity, which serves as a good inital model for subsequent FWI. With a good initial subsalt velocity, FWI can reconstruct the model velocity even with the absence of low frequencies or long offsets.    
\\
\\
Most of the cost of the training is in generating the training set, and specifically in the inversion of the 1D models. We used deepwave \citep{deepwave} to implement a fast FWI inversion on GPUs. We utilized four NVIDIA Tesla V100 GPU cards, and in each card, we perform FWI on 2000 models. The estimated time to generate the training set was about three hours. However, the training itself of the neural network lasted about thirty minutes using only one GPU card. Considering that the training samples generation and the NN training cost is the same whether we have 2D or 3D models, the overall cost of training the network can be considered small compared to the cost of applying FWI on the whole data.
\section*{Conclusion}
We proposed to automatically unflood the salt using deep learning, specifically a U-net architecture. The network is trained in a regression manner using velocity models obtained by FWI, unlike the usual salt detection networks that are trained as classifiers on seismic images. Training directly on the velocity with a regression regime enabled the network to detect the \textbf{BoS} and estimate the subsalt velocity. For training, we created many 1D salt and non-salt models and performed FWI on them after flooding the salt. The inputs to the network are the results from FWI as well as the corresponding initial models and the outputs are the true velocity models. 
\\
\\
We consider three scenarios in generating the training datasets. In the first scenario, the generated data does not contain any prior geological knowledge about the model of interest that we want to apply the network on, the BP model in our example. This results in a mismatch in the statistics of the generated data and the BP model leading to poor performance, especially in estimating the subsalt velocity. The second experiment includes partial knowledge of the BP model, such as the maximum velocity and having similar water depth. The data in the second experiment show similar statistics with BP in the deeper parts; thus, the network unfloods the salt with a fair estimate of the subsalt velocity. The data in the third experiment was created using the average velocity of the BP sediments, excluding the salt as a background. In real life, this could be somewhat replicated by taking the average of the velocities measured at different well locations. The basic statistics of the generated data and the model of interest, in this case, are similar. The resulting unflooding showed a homogeneous estimation of the subsalt velocity that is around the average velocity of BP at that depth. We showed that when the method succeeds in estimating the subsalt velocity, a follow-up FWI can provide a good subsalt model even in the absence of low frequencies and sufficient offsets. This is important as many vintage seismic data do not have the frequencies and offsets needed for a successful FWI.

\bibliographystyle{chicago}  
\bibliography{ref}

\end{document}